\documentclass[epj]{svjour}
\usepackage{graphics}
\begin{document}
\title{Regularization of singular terms in $N\overline{N}$
potential model}

\author{O.D. Dalkarov\inst{1}\and A. Yu. Voronin\inst{1}
}                     
%
%
\institute{P. N. Lebedev Physical Institute\\53 Leninsky pr.,117924
Moscow, Russia}
\date{Received: date / Revised version: date}
%
\abstract{ We suggest a method of singular terms regularization  in
potential model of $N\overline{N}$ interaction. This method is free
from any uncertainties, related to the usual cut-off procedure and
based on the fact, that in the presence of sufficiently strong
short-range annihilation $N$ and $\bar{N}$ never approach close
enough to each other. The effect of mentioned singular terms of OBE
potential, modified by annihilation is shown to be repulsive. The
obtained results for S- and P-wave scattering lengths are in
agreement with existing theoretical models.
} 
\maketitle
\section{Introduction}
\label{intro}

During the last decades numerous nonrelativistic models of
$N\overline{N}$ low energy interaction
\cite{Phil,Shapiro,DM,TU,Par82,CPS,KW,DR,Pign,Par99} have been
suggested. An intriguing problem of possible existence of the so
called quasi-nuclear $N\overline{N}$ states \cite{Shapiro} strongly
stimulated the mentioned researches. The physical arguments in favor
of such  states are the following. The interaction between $N$ and
$\overline{N}$ should be much more attractive than the $NN$
interaction, as it follows from the procedure of $G$-conjugation
\cite{Phil}. Such a strong attraction should produce a spectrum of
$N\overline{N}$ quasi-bound states (so called \textit{baryonium}).
In the same time the range of annihilation, estimated from the
position of the nearest to the threshold singularity in the Feynman
annihilation diagrams, is much smaller than the range of the meson
exchange forces. This means that baryonium states could be rather
narrow to be observed experimentally. It was indeed  discovered in
the Low Energy Antiproton Ring (LEAR) experiments
\cite{LEAR1,LEAR2,LEAR3} that certain partial cross-sections sharply
increase with decreasing of energy down to the $N\overline{N}$
threshold (so called $P$-wave enhancement), which could be a
manifestation of the narrow weakly-bound state or resonance. This
conclusion
 was verified by the experiments with antiprotonic atoms
\cite{Gotta,At1} and detailed experimental studies  of antiproton
annihilation at rest by OBELIX collaboration \cite{OB1,OB2,OB3,OB4}.
The recent experimental data \cite{JPsi} on $J/\psi$ decay into
$\gamma p \bar{p}$ channel also indicate a strong enhancement near
the $p\bar{p}$ threshold.

However, the transparent physical picture of the quasi-nuclear
states has a significant drawback. The $G$-conjugation of $NN$ OBE
potential yields in attractive singular terms in $N\overline{N}$
potential of the type $1/r^{3}$. (In case of $NN$ these terms are
repulsive and play a role of the so called short range core). It is
well known, that attractive singular potentials produce a collapse
\cite{LL}, i.e. the spectrum of the system is not bounded from
below, while the scattering problem has no definite solution. The
usual way of dealing with such pathological potentials is to impose,
that singular behavior is an artifact of certain approximations (for
instance nonrelativistic approximation). In the absence of the
self-consistent theory it is common practice to introduce the
phenomenological cut-off radius to regularize the singular behavior
of the model at short distance. However the results change
dramatically with small variations of the cut-off radius (as long as
we deal with a real part of $N\overline{N}$ potential) \cite{Carb}
and depend on the details of the cut-off procedure, which seriously
diminish the predictive power of the model. In fact it is not clear
if the near-threshold states are determined by the ''physical part''
of the OBEP, or they are artifacts, produced by the
''non-physical'', singular part of the interaction.

The aim of the present study is to analyze the role of the singular
terms and suggest a model of $N\overline{N}$ interaction which is
free from the mentioned above uncertainties of the cut-off
procedure. The main idea of our approach is that strong enough short
range annihilation makes low energy scattering observables
independent on any details of the short-range interaction, as far as
the particle annihilates rather than "falls to the center".

 Mathematically this means  that attractive singular potential
 becomes \emph{regular}  when gets the \textit{imaginary }addition to the
 interaction strength \cite{AV}. Curiously this is true even if such
 an
addition is infinitesimal. It is shown that the scattering on the
regularized in such a way singular potential is equivalent to the
full absorption of the particles in the scattering center.
Encouraged by the early result of Dalkarov and Myhrer \cite{DM}, who
introduced a
  full absorption boundary condition at certain inter-baryonic distance
  and successfully described the low energy $N\bar{N}$
  scattering data,  we suggest a regular potential model of $N\overline{N}$ interaction,
based on OBE potential, but without any cut-off radius.

The important feature of our model is that the reflected wave is
generated only by those parts of OBE potential, where WKB
approximation fails, i.e. by the medium and the long range part of
the OBEP. This means that any information about the low energy
$N\overline{N}$ scattering could be determined by the mentioned
parts of OBEP only.
 In particular we will show that  ''singular''
part of the potential, modified by  annihilation  cannot produce any
quasi-bound states.  The near-threshold resonances, which are well
reproduced by our model, are determined by the long range part of
OBEP. We calculate the scattering lengths in S- and P- partial waves
for different values of spin, isospin and total momentum and
demonstrate that obtained results are in good agreement with
existing theoretical models \cite{KW,DR}

The paper is organized as follows. In the second section we discuss
the general properties of attractive singular potentials,
regularized by imaginary addition to the interaction strength. The
third section is devoted to the application of the developed
approach to the construction of the regularized $N\overline{N}$
potential model.

\section{Inverse power potentials with complex strength.}

  In this section we present the main results concerning the
properties of homogeneous potentials $-\alpha _{s}/r^s$ with a
complex strength $\alpha _{s}=\mathop{\rm Re}\alpha _{s}\pm i\omega
$. In the following we put $2M=1$. Let us first treat the case
$s>2$. Near the origin one can neglect all the  terms of the
Shr\"{o}dinger equation, increasing  slower than $1/r^{2}$ and get
the following expression for the wave-function
 \cite{MM}:

\begin{eqnarray}
\Phi (r) &=&\sqrt{r}\left(H_{\mu }^{(1)}(z)+\exp
(2i\delta_0)H_{\mu}^{(2)}(z)\right)\label{Phi}\\
z &=&\frac{2\sqrt{\alpha _{s}}}{s-2}r^{-(s-2)/2}\\
\mu &=&\frac{2l+1}{s-2}
\end{eqnarray}

Here $H_{\mu }^{(1)}(z)$ and $H_{\mu }^{(2)}(z)$ are the Hankel
functions of order $\mu $ \cite{Watson}, $\delta_0$ is a
contribution of the short range part of the inverse power potential
into the scattering phase. It is worth to mention that the variable
$z$ is a semiclassical phase.

Let us replace the inverse power potential at distance less than
$r_{0}$ by the constant  $-\alpha_s/r_{0}^s$, having in mind to tend
$r_0\rightarrow 0$. Matching the logarithmic derivatives for the
"square-well" solution and the solution (\ref{Phi}) at small $r_0$ ,
one can get for $\delta_0$:
\begin{eqnarray}
\delta_0&=&p(r_0)r_0 \label{delta0} \\
p(r_0)&=&\frac{\sqrt{\mathop{\rm Re}\alpha _{s}\pm i\omega
}}{r_0^{s/2}}\label{peff}
\end{eqnarray}

Now it is important that the interaction strength $\alpha_s$ is
complex. In the limit $r_0\rightarrow 0$ we obtain:

\begin{equation}
\lim_{r_0\rightarrow 0}\mathop{\rm Im}\delta _{0}=\mathop{\rm
Im}\frac{\sqrt{\mathop{\rm Re}\alpha _{s}\pm i\omega
}}{r_0^{(s-2)/2}}\rightarrow \pm \infty \label{Limdelt}
\end{equation}
which means, that $\exp(2i\delta_0)$ is either $0$ or $\infty$ and
the linear combination of the Shr\"{o}dinger equation solutions
(\ref{Phi}) is uniquely defined in the limit of zero cut-off radius
$r_0$:
\begin{equation}
\lim_{r_0\rightarrow 0}\Phi(r)=\left\{
\begin{array}{cll}
\sqrt{r}H_{\mu }^{(1)}(z) & \mbox{if} & \omega > 0 \\
\sqrt{r}H_{\mu }^{(2)}(z) & \mbox{if} & \omega < 0%
\end{array}
\right.  \label{bound}
\end{equation}
One can see, that $\omega > 0$ selects an incoming wave , which
corresponds to the full absorption of the particle in the scattering
center, while $\omega < 0$ selects an outgoing wave, which
corresponds to the creation of the particle in the scattering
center.

As one can see from (\ref{delta0}) and (\ref{peff}) as long as
$s>2$, the above conclusions are valid for any infinitesimal value
of $\omega$. It means, that the sign of an infinitesimal imaginary
addition to the interaction constant selects the full absorption or
the full creation boundary condition (\ref{bound}). This boundary
condition can be formulated in terms of the logarithmic derivative
in the origin:
\begin{equation}
\lim_{r\rightarrow 0}\frac{\Phi'(r)}{\Phi(r)}=- i \mathop{\rm
sign}(\omega) p(r)
\end{equation}
where $p(r)$ is a classical local momentum (\ref{peff}). (Compare
with plane incoming (outgoing) wave boundary condition $\exp(\mp
ipr)'/\exp(\mp ipr)=\mp i p$).

As soon as the solution of the Shr\"{o}dinger equation is uniquely
defined, we can  calculate the scattering observables. In particular
we can now obtain the S-wave scattering length for the potential
$-(\alpha _{s}\pm i0)/r^{s}$ (for $s>3$) :

\begin{equation}
a=\exp (\mp i\pi /(s-2))\left( \frac{\sqrt{\alpha _{s}}}{s-2}\right)
^{2/(s-2)}\frac{\Gamma ((s-3)/(s-2))}{\Gamma
((s-1)/(s-2))}\label{sclength}
\end{equation}

The fact, that in spite $\mathop{\rm Im}\alpha _{s}\rightarrow \pm
0$ the scattering length has nonzero imaginary part is the
manifestation of the singular properties of attractive real inverse
power potential with $s>2$ which violates the self-adjointness of
the Hamiltonian.

 Let us compare the scattering length (\ref%
{sclength}) with that of the repulsive inverse power potential $\alpha _{s}/r^{s}$%
. One can get:
\begin{equation}
a^{rep}=\left( \frac{\sqrt{\alpha _{s}}}{s-2}\right)
^{2/(s-2)}\frac{\Gamma ((s-3)/(s-2))}{\Gamma ((s-1)/(s-2))}
\label{screpulsive}
\end{equation}

It is easy to see, that (\ref{sclength}) can be obtained from (\ref%
{screpulsive}) simply by choosing the certain branch (corresponding
to an absorption or a creation) of the function $\left( \sqrt{\alpha
_{s}}\right) ^{2/(s-2)}$ when passing through the branching point
$\alpha _{s}=0$. The scattering length in a regularized inverse
power potential becomes an analytical function of $\alpha _{s}$ in
the whole complex  plane of $\alpha _{s}$ with a cut along positive
real axis. One can see, that the presence of an inelastic component
in the inverse power potential acts in the same way, as a repulsion.
It suppresses one of two solutions of the Schr\"{o}dinger equation
and thus eliminates the collapse.

It is easy to see, that the boundary condition (\ref{bound}) of the
full absorption (creation) is incompatible with the existence of any
bound state. Indeed, one needs both incoming and reflected wave to
form a standing wave, corresponding to a bound state. This means,
that the regularized inverse power potential supports \textbf{no
bound states}. This is also clear from the mentioned above fact,
that the scattering length for a regularized attractive inverse
power potential is an analytical continuation of the scattering
length of a repulsive potential.


Let us now turn to the very important case $-\alpha /r^{2}$. The
wave-function now is:
\begin{eqnarray}
\Phi  &=&\sqrt{r}\left[ J_{\nu _{+}}(kr)+\exp
(2i\delta_0)J_{\nu _{-}}(kr)\right]  \\
\nu _{\pm } &=&\pm \sqrt{1/4-\alpha _{2}}
\end{eqnarray}
where $k=\sqrt{E}$, and $J_{\nu _{\pm }}$ are the Bessel functions \cite%
{Watson}. In the following we will be interested in the values of
$\mathop{\rm Re}\alpha _{2}$ greater than critical $\mathop{\rm
Re}\alpha _{2}>1/4$. We use the same cut-off procedure at small
$r_0$.
Matching the logarithmic derivatives at cut-off point $r_0$ we get for $%
\exp (2i\delta_0)$:
\[
\lim_{r_0\rightarrow 0}\exp (2i\delta_0)=r_{0}^{\nu _{+}-\nu
_{-}}const\sim r_{0}^{2\nu }=r_{0}^{\omega /\sqrt{\mathop{\rm
Re}\alpha _{2}-1/4}}r_{0}^{-2i\sqrt{\mathop{\rm Re}\alpha _{2}-1/4}}
\]

One can see, that due to the presence of an imaginary addition $
\omega$ we get $\mathop{\rm Im}\delta _{0}\rightarrow \pm \infty$
when $r_{0}\rightarrow 0$.

Again we come to the boundary condition:
\begin{equation}
\lim_{r_0\rightarrow 0}\Phi(r)=\left\{
\begin{array}{cll}
\sqrt{r}J_{\nu _{+}}(kr) & \mbox{if} & \omega > 0 \\
\sqrt{r}J_{\nu _{-}}(kr) & \mbox{if} & \omega < 0%
\end{array}
\right.  \label{bound2}
\end{equation}
where $\nu _{_{\pm}}=\pm\sqrt{1/4-\alpha _{2}}$

For the large argument this function behaves like:
\[
\Phi \sim \cos (z-\nu _{_{\pm}}\pi /2-\pi /4)
\]
The corresponding scattering phase is:
\begin{equation}
\delta =\frac{\pm i\pi }{2}\sqrt{\alpha _{2}-1/4}+\pi /4
\label{scphase}
\end{equation}

As one can see, the S-matrix $S=\exp(2i\delta)$ is energy
independent. This means that the regularized inverse square
potential supports \textbf{no bound states}. The regularized
wave-function and the phase-shift are analytical functions of
$\alpha _{2}$ in the whole complex plane with a cut along the axis
$\mathop{\rm Re}\alpha _{2}>1/4.$

\subsection{Critical strength of inelastic interaction}
Now we would like to determine "how strong" should be annihilation
to regularize the real attractive singular potential of order $s$.
In other words we would like to find the minimum power $t$ of an
\emph{infinitesimal} imaginary inverse power potential required for
the regularization of given singular potential. The potential of
interest is a sum $-\alpha _{s}/r^{s}\mp i\omega/r^{t}$. Here we
keep $\alpha _{s}$ real. From expressions (\ref{delta0}, \ref{peff})
one immediately comes to the conclusion that the regularization is
possible only if:
\[
t>s/2+1
\]

 Thus we come to the conclusion that the scattering is insensitive to
 any details of the short range modification
 of a
singular interaction if the inelastic component of such an
interaction increases in the origin faster than $-1/r^{(s/2+1)}$.

\subsection{Singular potential and WKB approximation.}
The WKB approximation  holds if $|\frac{\partial (1/p)}{\partial
r}|\ll 1$. In case of the zero-energy scattering on a regularized
inverse power potential with $s>2$ this condition is valid for:
\[ r\ll
r_{sc}\equiv (2\sqrt{\alpha _{s}}/s)^{2/(s-2)}
\]

 (For $s=2$ the semiclassical approximation is valid only
for $\alpha_2 \gg 1$).
The WKB approximation, consistent with the boundary condition (\ref{bound}) for $%
s>2 $ is :
\begin{equation}
\Phi =\frac{1}{\sqrt{p(r)}}\exp (\pm i\int\limits_{r}^{a}p(r)dr)
\label{semicl}
\end{equation}
with $p(r)$ from (\ref{peff}). It follows from the above expression,
that in case the WKB approximation is valid everywhere the solution
of the Schr\"{o}dinger equation  includes incoming wave only (for
distinctness we speak here of absorptive potential). The
corresponding S-matrix is equal to zero $S=0$ within such an
approximation and insensitive to any details of the inner part of
potential $p^2(r)$. The outgoing wave can appear in the solution
only in the regions where (\ref{semicl}) does not hold. For example,
in the zero energy limit $k^2\rightarrow 0$ the S-matrix is nonzero
$S=1-2ika$. One can show that the outgoing wave is reflected from
those parts of the potential which change sufficiently fast in
comparison with the effective wavelength (so called quantum
reflection) $|\frac{\partial (1/p)}{\partial r}|\\geq 1$

For the zero energy scattering and $l=0$ this holds for $r\geq
r_{sc}$.

The reflection coefficient $P\equiv|S|^2$ which shows the reflected
part of the flux  has the following form in the low energy limit:
\[
P=1-4 k\mathop{\rm Im}a
\]
For the energies $E\gg E_{sc}\equiv(s/2)^{2s/(s-2)}\alpha
_{s}^{-2/(s-2)}$ the WKB holds everywhere and S-wave reflection
becomes exponentially small.

 An important conclusion is that any
information, which comes from the scattering on an absorptive
singular potential is due to a quantum reflection from the tail of
such a potential.
\subsection{Near-threshold scattering and bound states}

Above we have shown that there are no bound states in purely
homogenies attractive inverse power potential with complex strength
(including the case of infinitesimal imaginary part). The physical
reason is the absence of the reflected wave from the absorptive core
of inverse power potential. In this subsection we will be interested
how the low energy scattering amplitude and spectrum of the
near-threshold states of a regular potential $U(r)$ is modified by a
potential, which has inverse power behavior $-(\alpha
_{s}+i0)/r^{s}-\alpha _{2}/r^{2}$ near the origin. The cases when
such a modification is small are of our special interest.

\subsubsection{Penetration under the centrifugal barrier}
One could expect that  when regularized singular interaction is
separated from the regular one by the centrifugal barrier the effect
of the regularized singular terms could be small and determined by
the
 centrifugal barrier penetration probability.
 In fact, if $\alpha _{s}$ is small enough, there is a range where
\begin{equation}
U(r)\ll \alpha _{s}/r^{s}\ll (l(l+1)-\alpha _{2})/r^{2}
\label{domain}
\end{equation}%
Let us suggest that the regular potential is approximately
constant in the mentioned range of $r$, so that $U(r)\approx p^{2}$. Then from (\ref{domain}%
) we get:

\begin{equation}
p\alpha _{s}^{1/(s-2)}\ll 1  \label{pro}
\end{equation}

For such values of $r$ the wave function is:

\begin{eqnarray}
&&\Phi \sim \sqrt{r}(J_{\mu }(pr)-\tan (\delta _{s})Y_{\mu }(pr))
\label{Free} \\
Y_{\mu } &=&\frac{J_{\mu }\cos (\mu \pi )-J_{-\mu }}{\sin (\mu \pi )} \\
&&\mu =\sqrt{(l+1/2)^{2}-\alpha _{2}}  \nonumber
\end{eqnarray}

here $\delta _{s}$ is a phase shift, produced by the regularized
singular potential in the presence of the regular potential.

For small $r\sim \alpha _{s}^{1/(s-2)}$ the wave-function is
determined by the regularized singular and centrifugal potential:

\begin{eqnarray*}
\Phi &\sim &\sqrt{r}H_{\nu }^{+}(\frac{2\sqrt{\alpha_{s}
}}{s-2}r^{-(s-2)/2})
\\
\nu &=&2\mu /(s-2)
\end{eqnarray*}

Matching the logarithmic derivatives and taking into account
(\ref{pro}) we get for the phase shift $\delta _{s}$:

\[
\delta _{s}=-\gamma_{\mu} (\frac{p\alpha _{s}^{1/(s-2)}}{2(s-2)^{2/(s-2)}}%
)^{2\mu }\exp (-i\pi \nu )
\]
where \[ \gamma_{\mu}= \sin (\pi \mu )\frac{\Gamma (1-\mu )\Gamma
(1-\nu )}{\Gamma (1+\mu )\Gamma (1+\nu )}\]

Let us mention, that for nonzero $l$ the above expression for
$\mathop{\rm Re}\delta _{s}$ may become inaccurate, as far as
 the  phase shift, produced by
the tail of the regularized singular potential, may become greater
than the phase shift (\ref{delta}), produced by the core of the
regularized singular potential. Such a correction depends on the
certain form of the tail of the regularized singular potential and
can be calculated as a first order of a distorted wave
approximation.

In the same time $\delta _{s}$ has positive imaginary part according
to the "inelastic" character of regularized singular potential.

\begin{equation}
\mathop{\rm Im}\delta _{s}=(-1)^{l}(\frac{p\alpha _{s}^{1/(s-2)}}{2(s-2)^{2/(s-2)}}%
)^{2l+1}\gamma_{\mu}  \label{Imdelta}
\end{equation}

The near-threshold states produced by the regular potential $U(r)$
are perturbed by the short range regularized singular potential. In
particular they get the widths,
which in our case of small $\delta _{s}$ are proportional to $\mathop{\rm Im}%
\delta _{s}$.

If the near-threshold states spectrum in $U(r)$ has a semiclassical
character, than from the quantization rule:

\[
\int \sqrt{E_{n}+\delta E_{n}-U\left( r\right) }dr+\delta _{s}=const
\]

one gets:

\begin{equation}
\delta E_{n}=-\delta _{s}\omega _{n}  \nonumber
\end{equation}

where $\omega _{n}$ is a semiclassical frequency:

\[
\omega _{n}=(\int (E_{n}-U\left( r\right) )^{-1/2}dr)^{-1}
\]

Taking into account (\ref{Imdelta}) we get for the width of the
state:

\[
\Gamma _{n}/2=(-1)^{l+1}(\frac{p\alpha _{s}^{1/(s-2)}}{2(s-2)^{2/(s-2)}}%
)^{2l+1}\gamma_{\mu}\omega _{n}
\]

Thus in the above mentioned case the modification of the
near-threshold spectrum of the regular potential $U=p(r)^2$ by the
regularized singular potential results in shifting and inelastic
broadening determined by the small parameter $(p\alpha
_{s}^{1/(s-2)})^{2l+1}$, which characterizes the centrifugal barrier
penetration probability.

\subsubsection{ Quantum reflection states}

 We will treat
here an interesting case, when there is no barrier separation
between (absorptive) regularized singular and regular parts of
interaction. However the existence of the narrow near-threshold
states is still possible. The reason why in such a case rather
narrow states can survive is the so called quantum (over-barrier)
reflection from those parts of \textit{attractive} potential, which
change sufficiently fast.

To illustrate this idea let us expect that the full interaction
potential $U(r)$ has the following form:
\begin{equation}
U(r)=\left\{
\begin{array}{cll}
-(\alpha _{s}+i0)/r^{s} & \mbox{if} & r<r_{0} \\
-(\alpha _{s}/r_{0}^{s})\Theta(r-R) & \mbox{if} & r\geq r_{0}%
\end{array}
\right.  \label{QRpot}
\end{equation}

This potential can be treated as a shallow and wide square-well with
depth $\alpha _{s}/r_{0}^{s}$ and width $R$ perturbed by the
regularized singular interaction $-(\alpha _{s}+i0)/r^{s}$, which is
cut at distance $r_0$. We are interested in the behavior of the
square-well near-threshold spectrum under such a perturbation. For
the moment we will restrict our treatment with the S-wave case only.

 Let us choose $r_0>
r_{sc}\equiv (2\sqrt{\alpha _{s}}/s)^{2/(s-2)}$. It was shown above
that the WKB approximation, applied to the zero energy scattering on
the regularized singular potential fails for $r>r_{sc}$. Thus in our
problem there is a domain  $r_{sc}<r \leq r_0$ of WKB failure. We
will show that this domain acts similar to the barrier in the sense
that it is responsible for the reflected wave generation.

 For
$r\sim r_0$ the zero energy wave-function in the regularized
singular potential has the form:
\[
\Phi(r)\sim 1-r/a
\]
Here $a$ is the \textit{complex} scattering length (\ref{sclength})
in the regularized singular potential $-(\alpha _{s}+i0)/r^{s}$.
 The wave-function in the square well is:
 \[
 \Phi(r)\sim \sin(kr+\delta)
 \]
 Here $k^2=\alpha _{s}/r_{0}^{s}-E$, where $E$ is the energy of the
 near-threshold state, while $\delta$ is the phase-shift, produced
 by the regularized singular potential.
 Matching the logarithmic derivatives at point $r_0$ and expecting that $|kr_0+\delta| \ll 1$ we find:
 \[
 \delta=-ka
 \]

 To ensure the existence of the
near-threshold state of interest (such that $|k(r_0-a)|\ll 1$)  we
choose $R$ big enough. The required characteristic value $R_c$ can
be obtained from the condition of the state appearance in the square
well of the depth $\alpha _{s}/r_{0}^{s}$:
\[
\sqrt{\alpha _{s}/r_{0}^{s}}R_c=\pi/2
\]

Matching the logarithmic derivatives of the square well
wave-function and the decaying wave at point $R$ we get for the
bound energy $E=-\kappa^2$:
\[
k\cot(k(R-a))=-\kappa
\]
which for $\kappa\ll k$  gives:
\begin{eqnarray}
k&\simeq & \sqrt{\alpha _{s}/r_{0}^{s}}\\
\mathop{\rm Re}\kappa&=&k^2(R-R_c-\mathop{\rm
Re}a)\\
 \mathop{\rm Im}\kappa&=-&k^2\mathop{\rm
Im}a
\end{eqnarray}

The corresponding S-matrix pole in the complex $k-$plane is
$z=i\kappa$. One can see, that depending on the sign of $\mathop{\rm
Re}\kappa$ the mentioned pole can be either in the upper half-
plane, which corresponds to the bound state, or in the lower
half-plane, which corresponds to the virtual state. The width of the
state is proportional to the imaginary part of the regularized
singular potential scattering length:
\[\Gamma/2=-k^4\mathop{\rm
Im}a(R-R_c-\mathop{\rm Re}a)
\]

One can see, that the effect of the regularized singular potential
is determined by the parameter $k\mathop{\rm Im}a$. The physical
sense of such a parameter can be easily established. In fact, the
S-matrix element corresponding to the
 scattering with a small momentum $k$ on the regularized singular potential is:
\[
S=1-2ika
\]
The intensity of the reflected wave is:
\[
|S|^2=1-4k|\mathop{\rm Im}a|
\]

The smaller is $k\mathop{\rm Im}a$ the higher is the probability of
quantum reflection and less is the influence of the regularized
singular potential on the near-threshold spectrum of the  regular
potential.

One can see, that  the mentioned above reflection takes place from
the domain of WKB failure. ( Indeed, as it was shown above in case
WKB is valid everywhere the full absorbtion  in the origin would
ensure that there is no reflected wave, i.e. S-matrix is zero.) This
phenomenon of quantum (over-barrier) reflection is known for a long
time \cite{LJ} in different fields, such as neutron physics or
physics of ultra-cold atomic collisions \cite{QR}.

From the above treatment one can get the estimation for the maximum
binding energy in  a regular potential $U(r)$ modified by an
absorptive regularized singular potential.

Let $r_0$ is the distance where the WKB failure takes place. As we
have shown the effect of the WKB failure domain  is the partial
reflection. Thus, for the purpose of the qualitative estimation, we
can replace this domain by the boundary condition of full reflection
at $r_0$. In other words one should look for the bound states in the
following truncated potential:
\begin{equation}
U_{tr}(r)=\left\{
\begin{array}{cll}
+\infty & \mbox{if} & r<r_{0} \\
U(r) & \mbox{if} & r\geq r_{0}%
\end{array}
\right.  \label{Utr}
\end{equation}
 The ground state energy $E_{tr}$ in such a potential gives an
 approximation for the lowest (quasi-)bound state energy in the full
 potential $U(r)$.

\section{Optical model of $N\overline{N}$ interaction.}

From the above results it is clear that the model potential, which
behaves at short distance like $-(\alpha+i\omega)/r^{3}$ is regular,
i.e. it enables definite unique solution of the scattering problem.
Such a potential is absorptive and describes not only elastic, but
inelastic scattering as well. The above statements are true even for
infinitesimal value of $\omega$. As we have shown such a
infinitesimal imaginary addition is equivalent to the full
absorption boundary condition in the origin. We apply this
regularization procedure to the $N\overline{N}$ potential in
$^{13}P_{0}$ state. We use the version of real OBEP potential  from
the Kohno-Weise model \cite{KW} accompanied by  the imaginary
component $-i\omega /r^{3}$ with $\omega \rightarrow 0$:
\[
W=W_{OBEP}-i\omega /r^{3}
\]
 Here $W_{OBEP}$ is the Kohno-Weise real potential \textbf{without any
   cut-off.}

The scattering volume $^{3}P_{0}$ $T=0$ calculated in the limit
$\omega \rightarrow 0$ turns to be $a_{r}=-7.66-i4.87$ fm$^3$, while
the value  obtained within Kohno-Weise model with a cut-off
$r_{c}=1$fm is $a_{KW}=-8.83-i4.45$ fm$^3$. As one can see, both
scattering volumes are rather close. This procedure can be
successfully applied to any $N\overline{N}$ state, including
attractive singular potential terms.

Obviously, the $N\overline{N}$ states where singular attractive
terms are absent do not need any  regularization. In this case the
inelastic processes can be described by regular imaginary potential,
which parameters are carefully fitted for each partial wave. However
it is desirable to check if the real OBE potential accompanied with
the model of full absorption of the particles in a small volume
around the origin can describe the whole set of low energy
$N\overline{N}$ scattering data.

We suggest the following modification of the $N\overline{N}$
potential model:
\begin{equation}
W=V_{KW}-i\frac{A}{r^{3}}\exp(-r/\tau)\label{MP}%
\end{equation}
here $V_{KW}$ is the real part of Kohno-Weise version of OBE
potential \cite{KW}, but \textit{without any cut-off}. The
parameters of the imaginary part of the potential were choused as
follows: A=4.7 $GeV$ $fm^{2}$, $\tau=0.4$ $fm$.  We have calculated
the values of S- and P-scattering lengths in such a model potential.
The obtained results, (indicated as Reg) together with the results
of two Dover-Richard models, (DR1 and DR2), and Kohno-Weise model
(KW) taken from \cite{Carb1} are presented in the  Table I. In this
table the S-wave scattering lengths are given in $fm$, while the
P-wave scattering volumes are given in $fm^3$.

One can see rather good agreement between the results obtained
within the suggested optical model without cut-off and the cited
above versions of Kohno-Weise and Dover-Richard models.

\begin{table}
\centering
\begin{tabular}{|c|c|c|c|c|}
\hline
 State & DR1 & DR2 & KW & Reg\\ \hline
$^{11}S_{0}$ & 0.02-i1.12 & 0.1-i1.06 & -0.03-i1.35 & -0.08-i1.16\\
\hline $^{31}S_{0}$ & 1.17-i0.51 & 1.2-i0.57 & 1.07-i0.62 &
1.05-i0.55\\ \hline $^{13}S_{1}$ & 1.16-i0.46 & 1.16-i0.47 &
1.24-i0.63 & 1.19-i0.64\\ \hline $^{33}S_{1}$ & 0.86-i0.63 &
0.87-i0.67 & 0.71-i0.76 & 0.7-i0.65\\ \hline $^{11}P_{1}$ &
-3.33-i0.56 & -3.28-i0.78 & -3.36-i0.62 & -3.19-i0.59\\ \hline
$^{31}P_{1}$ & 0.92-i0.5 & 1.02-i0.43 & 0.71-i0.47 & 0.81-i0.46\\
\hline $^{13}P_{0}$ & -9.58-i5.2 & -8.53-i3.51 & -8.83-i4.45 &
-7.67-i4.74\\ \hline $^{33}P_{0}$ & 2.69-i0.13 & 2.67-i0.15 &
2.43-i0.11 & 2.46-i0.15\\ \hline
$^{13}P_{1}$ & 5.16-i0.08 & 5.14-i0.09 & 4.73-i0.08 & 4.75-i0.15\\
\hline $^{33}P_{1}$ & -2.08-i0.86 & -2.02-i0.7 & -2.17-i0.95 &
-2.09-i0.79\\ \hline $^{13}P_{2}$ & 0.04-i0.57 & 0.22-i0.56 &
-0.03-i0.88 & -0.12-i0.82\\ \hline $^{33}P_{2}$ & -0.1-i0.46 &
0.05-i0.55 & -0.25-i0.39 & -0.14-i0.39\\ \hline
\end{tabular}
\caption{S- and P-wave scattering lengths}
\end{table}
\bigskip

Let us outline here, that there are no reasons to believe that the
physical interaction indeed has the above form at small distances.
The physical sense of suggested strong annihilation model is that
the low energy scattering observables are independent \emph{on any
certain form} of short range interaction, and can be obtained by the
solution of the Shr\"{o}dinger equation which is \emph{formally}
applied to short distances.

\subsection{Near-threshold resonances}
The critical question of the quasi-nuclear model is weather the
strong annihilation could be compatible with the existence of narrow
quasi-bound $N\bar{N}$ states. It follows from the above treatment
of our regularized potential model that only few (if any)
near-threshold quasi-bound states or resonances can survive, while
any deeply bound states are excluded.

We examined the S-matrix poles in the state with the quantum numbers
$J=0$, $S=1$, $L=1$, $T=0$. The real and imaginary part of the
scattering volume in such a state as they appear in our calculations
are very large $a=-7.67-i4.74$ $fm^3$.  Such a big value of the
scattering volume could be an indication of the near-threshold state
or resonance.

Indeed we found that the nearest to the threshold S-matrix poles are
situated in the third and fourth quadrant of the complex k-plane:

\[
k_{+}=44.8-i54.3 \mbox{ MeV/c  }  k_{-}=-58.4-i73.8 \mbox{ MeV/c}
\]

 In the absence of annihilation
such poles should be symmetrical with respect to the imaginary axis
and correspond to the near-threshold resonance. The short-range
annihilation brakes the left-right symmetry between such poles. Such
near-threshold poles will manifest themselves by a rapid increasing
of related amplitude and cross-section with the decreasing of energy
down to the threshold.

The found resonance belongs to the mentioned above "quantum
reflection" case. Indeed, the analysis of our model $N\bar{N}$
potential $U(r)$ (into which we include the centrifugal potential)
in the $^{13}P_0$ state shows that the centrifugal barrier  is
overcome by the attractive singular terms. Thus there is no barrier
between regularized singular absorptive core and the "regular" part
of the interaction. In the same time one can check that the WKB
failure condition:
\[
\partial/\partial r(\lambda_{DB})\geq 1
\]
takes place for $r \geq r_0= 1.4$ fm.
 Here $\lambda_{DB}= 1/\sqrt{2MU(r)})$ is the local de
Broglie wavelength in the case of the zero energy scattering. To
judge about the spectrum of $U(r)$ we treat, according to the
presented above approach, the potential  truncated at point
$r_0=1.4$ fm (\ref{Utr}). One can check that such a potential
supports no bound states and the nearest to the threshold S-matrix
pole indeed corresponds to the resonance.

Thus the P-wave enhancement is explained in the presented model by
the existence of the near-threshold resonance. It should be
specially mentioned that no deep bound states with given quantum
numbers could exist in our model in spite of  strong $N\bar{N}$
attraction.

It is worth to mention that suggested here regularizing procedure
can be applied to any OBEP-based optical model of the
$N\overline{N}$ low energy interaction.

\section{Conclusion}
We have found that the scattering observables are insensitive to any
details of the short range interaction, if such an interaction
includes strong inelastic component and can be found from the
solution of the Shr\"{o}dinger equation, formally applied to short
distances. Mathematically this means that singular attractive terms
of $N\bar{N}$ potential can be regularized by an imaginary addition
to the interaction strength. We analyzed the main properties of such
a regularization. In particular it was shown that regularized in
mentioned way singular homogenies potential supports no bound
states. The low energy scattering amplitude on such a potential is
determined by the quantum scattering from the region, where WKB
approximation fails ( the potential tail). The mentioned formalism
was used to build an optical model of $N\bar{N}$ low energy
interaction free from uncertainty, related to the cut-off parameter.
The good agreement of the results, obtained within our
regularization method and within different $N\bar{N}$ interaction
models was established. We prove that  no deep quasi-bound states
are possible within our model, while the low energy scattering
observables are determined mainly by the long range part of OBEP.
The spectrum of narrow quasi-nuclear states is concentrated near the
threshold. It is argued that the existence of such states is
possible due to the phenomenon of quantum (over-barrier) reflection
from those parts of $N\bar{N}$ potential which change sufficiently
fast. These domains of WKB failure play a role of effective barrier
between regular part of interaction and absorptive singular core. In
particular, we demonstrate the existence of the near-threshold
resonance with quantum numbers $J=0$, $S=1$, $L=1$, $T=0$,
responsible for the P-wave enhancement.

\section{Acknowledgement}

The research was performed under support of Russian Foundation for
Basic Research grant 02-02-16809.

\end{document}